\documentclass{article}
\usepackage{geometry}
\usepackage{float}
\usepackage{subfig}
\usepackage{graphicx}
\usepackage[super,sort&compress,comma,numbers]{natbib}

\bibpunct{}{}{,}{s}{,}{,}
\usepackage{natmove}

\begin{document}

\markboth{Dilnoza B Amirkulova and Andrew D White}
{Combining Enhanced Sampling with Experiment Directed Simulation of the GYG peptide}

\title{Combining Enhanced Sampling with Experiment Directed Simulation of the GYG peptide}

\author{Dilnoza B Amirkulova and  Andrew D White}

\maketitle

\begin{abstract}

Experiment directed simulation is a technique to minimally bias molecular dynamics simulations to match experimentally observed results. The method improves accuracy but does not address the sampling problem of molecular dynamics simulations of large systems. This work combines experiment directed simulation with both the parallel-tempering and parallel-tempering well-tempered ensemble replica-exchange methods to enhance sampling of experiment directed simulations. These methods are demonstrated on the GYG tripeptide in explicit water. The collective variables biased by experiment directed simulation are chemical shifts, where the set-points are determined by NMR experiments. The results show that it is possible to enhance sampling with either parallel-tempering and parallel-tempering well-tempered ensemble in the experiment directed simulation method. This combination of methods provides a novel approach for both accurately and exhaustively simulating biological systems.

\end{abstract}
\section{Introduction}

Molecular simulations are connected  to experiments through ensemble properties, called observables. Often experimental and simulated values disagree, due to challenges such as limited accuracy of force-fields and discrepancies between how experiments can be represented in a simulation. This disagreement can be improved by adding an auxiliary biasing potential energy function\cite{Bonomi2017,Allison2017}. For example, a simulation can be improved by incorporating experimental results as structural restraints in simulations\cite{Cavalli2007}. Often these restraints are a type of harmonic energy penalty that brings conformational ensembles closer to the experimentally observed values \cite{Robustelli2010,Torda1990}. Visiting states that are outside structural restraints is unfavored with harmonic biases. This introduces some ambiguity in how strong these energetic restraints should be.

To address the ambiguity in using energetic restraints to match experimental data, a maximum entropy principle\cite{Jaynes1957} can be applied to find an energy bias that is minimal\cite{Boomsma2014,Roux2013c}. This bias acts only in the dimension of the collective variable which corresponds to the experimental observable\cite{Pitera2012}. Experiment directed simulation (EDS) is an implementation of maximum entropy biasing. EDS uses an adaptive biasing method to converge to the maximum entropy bias that achieves agreement between molecular simulation collective variables and experimental observables\cite{White2014}. This bias is also provably the smallest change that can be made to match experimental data \cite{Roux2013c,White2014,White2015}.

EDS utilizes a single replica, whereas most other maximum entropy methods utilize multiple replicas\cite{Bonomi2017}. Other biasing methods, such as replica-exchange chemical shift restrained molecular dynamics simulations\cite{Robustelli2010} and restrained-ensemble molecular dynamics simulation method\cite{RouxBenoitandIslam2013} that utilize multiple replicas require multiple replicas in order to match simulated results to some reference values. This increases the required computational effort, although there is an improvement in sampling due to the inherent increase of samples with the replicas. EDS, being a single-replica technique, provides no improvement to sampling. EDS has been applied to classical molecular dynamics\cite{White2014}, ab-initio molecular dynamics\cite{White2017}, and coarse-graining \cite{Dannenhoffer-Lafage2016}. EDS was able to give more accurate results on dynamic properties such as the self-diffusion coefficient of lithium in electrolyte solutions\cite{White2014}, and the proton-diffusivity in ab initio water\cite{White2017} as well.

EDS is a biasing method and this is designed to match the ensemble average of some collective variable to a certain reference value. EDS lets the underlying simulation determine the fluctuations and underlying distribution of the collective variable. Only the expected value is shifted to match the experimental value. Because EDS is only a method of simulation, it cannot correct or mitigate any error in the reference value, such as experimental error, human error, or instrumentation error.
If the experimental value is instead a distribution or becomes a distribution because a Bayesian inference approach with prior belief is desired, other techniques allow matching to that desired distribution like experiment directed metadynamics\cite{White2015} or metainference\cite{Bonomi2016a}. 

Aside from accuracy, another challenge in molecular dynamics simulations is sufficient sampling of phase space. The so-called ``sampling problem''\cite{Clarage1995} is especially true in solvated macromolecules, where systems can become trapped in energy minima. EDS introduces a bias in the potential to match experimental averages; it is not an enhanced sampling technique. Thus, there is a need to use an enhanced sampling techniques with EDS. Multiple enhanced sampling methods have been developed to alleviate this problem\cite{Bernardi2015}, often by biasing sampling to occur along a certain collective variable\cite{Darve2008}. This presents a complication with EDS if both an enhanced sampling technique and EDS are applied on the same set of collective variables. For example, metadynamics \cite{Laio2002} will add Gaussian bias to the potential energy with respect to a collective variable, thereby modifying the potential of mean force (PMF) of that CV. EDS cannot be applied to a collective variable with a biased PMF, because EDS requires canonical samples. For example, if metadynamics is applied to $\Phi$ dihedral angle and EDS is applied $\Psi$ dihedral angle, the $\Psi$ angle needs to be re-weighted to achieve canonical sampling due to bias on $\Phi$ and the inherent correlation between the $\Phi$ and $\Psi$. EDS needs the $\Psi$ variable to be canonically sampled. Hence, EDS and Metadynamics cannot be applied simultaneously in a simulation. The well-tempered ensemble method (WTE) provides an alternative approach because it biases the fluctuations in energy, and with an assumption of the underlying potential energy distribution, does not affect the canonical sampling of other CVs. WTE can be further combined with parallel-tempering replica-exchange to improve sampling further\cite{Hansmann1997,bittner2008,Bonomi2010}.

In this work we consider the parallel-tempering well-tempered ensemble (PT-WTE) combined with EDS\cite{Bonomi2010} to improve both sampling and accuracy on an example system. Like in parallel-tempering replica-exchange, $N$ replicas of the simulation system are run concurrently. The coldest replica is the system of interest, yet it can escape free energy minima by swapping configurations with higher temperatures. The goal of PT-WTE is to reduce the number of replicas while maintaining similar replica efficiency\cite{Bonomi2010}. It has been shown that PT-WTE metadynamics decreases the number of replicas from 100 to 10 while maintaining the similar energy landscape of solvated tryptophan-cage protein \cite{Deighan2012}. It has also been shown to improve speed of convergence to an accurate FES in model and real systems\cite{Bonomi2010,Deighan2012}. The technique can be applied to larger systems such as proteins to efficiently sample their large conformational space\cite{Spiwok2015}.
The tri-peptide glycine-tyrosine-glycine (GYG) was used as a model system to demonstrate an implementation. GYG is found in structural and transport proteins such as collagen\cite{Bella2006} and ion channels\cite{Dibb2003}. Either mutations or complete deletions of this sequence from proteins showed either complete or partial loss  of a function in those proteins \cite{Thiagarajan2008,So2001,Berneche2005}, so it has specific functional relevance. The experimental data utilized here is from backbone chemical shifts \cite{Platzer2014} of GYG peptide. These are used to assess accuracy of the simulation and the free energy surface along the $\Phi$-$\Psi$ dihedral angles is used to assess sampling. 

Metadynamics metainference\cite{Bonomi2016} and replica-averaged metadynamics\cite{Camilloni2013} are the nearest examples of the method proposed here in literature. Metadynamics metainferences is a combination of the enhanced sampling technique of metadynamics\cite{Laio2002} with the ability to match experimental data of metainference\cite{Bonomi2016}. Replica-averaged metadynamics is the combination of the replica-averaging approach to matching experimental data\cite{Camilloni2013} with the enhanced sampling of metadynamics. A distinguishing difference between our approach and these is that the biasing force from EDS can be calculated with an enhanced-sampling method first and then be applied to a second simulation without replicas. This capability allows dynamic properties like autocorrelation functions or diffusivities to be computed.

\section{Theory}

In the WTE method\cite{Bonomi2010}, Newton's equation of motion becomes: 

\begin{equation}
    m \ddot{\mathbf{R}} = -\frac{U(\mathbf{R})}{\partial \mathbf{R}} - \frac{\partial V(U(\mathbf{R}), t)}{\partial \mathbf{R}}
\end{equation}

\noindent
where $m$ is the mass vector, $\mathbf{R}$ is the position vector, $U(\mathbf{R})$ is the system potential energy and $V(\cdot)$ is a \textit{time-dependent} function of the instantaneous potential energy which biases the system. $V(\cdot)$ is defined with the following time-derivative:

\begin{equation}
    \dot{V}(U, t) = \omega e^{-V(U, t) / k_B \Delta T} \delta{}(U - U(t))
\end{equation}

\noindent
where $\omega$ and $\Delta T$ are tuned constants. The intuition of this equation is to add a Guassian at each time $t$ centered at $U(t)$, just like in the metadynamics method \cite{Laio2002,Barducci2008}. The key parameter in WTE is the bias factor, like in well-tempered metadynamics\cite{Barducci2008}, which is defined relative to system temperature and $\Delta T$ as $\gamma = (T + \Delta T)/T$

Simulating according to this bias will asymptotically lead to the following distribution of potential energy, assuming normality of the potential energy probability distribution\cite{Bonomi2010}:
\begin{equation}
    P(U)^{1 / \gamma} \propto e^{\left(U - \left<U\right>\right)^2 / 2\gamma \Delta U^2}
\end{equation}

\noindent
where $\left<U\right>$ and $\Delta U^2$ are the expected value and variance of the potential energy of the unbiased canonical sampling. This expression shows that under the WTE bias, the average energies are the same but the variance of distribution is increased by $\sqrt{\gamma}$. A bias factor of 1 thus leads to canonical sampling. Increasing the bias factor increase the fluctuations in potential energy.

EDS modifies the potential energy of a system with a maximum entropy linear bias in a collective variable\cite{Pitera2012}. After equilibration, EDS will have the following biased potential energy: 

\begin{equation}
    U^{'}(\mathbf{R}) = U(\mathbf{R}) + \alpha f(\mathbf{R})
\end{equation}

\noindent
where $\alpha$ is a Lagrange multiplier determined from an equilibration process described in White et al.\cite{White2014} and $f(\mathbf{R})$ is a differentiable collective variable. The $\alpha$ is selected such that $\left<f\right> = \hat{f}$, where $\hat{f}$ is a pre-determined value for what the expected value of the collective variable should be. In this work, we have experimentally determined NMR chemical shifts for $\hat{f}$. The limitation of EDS is that finding $\alpha$ requires calculating $\left<f\right>$ with good sampling of the underlying canonical distribution. The PT-WTE method can ensure we have sufficient sampling of the distribution of $f$ values. The WTE process only needs to be modified to replace $U(\mathbf{R})$ with $U^{'}(\mathbf{R})$ and the previous analysis holds.

The benefit of WTE is that the increased energy fluctuations arising from the choice of bias factor can improve replica-exchange efficiency in a parallel-tempering setting. The probability of exchange for replicas $i$ and $j$ becomes\cite{Bussi2014}:
\begin{equation}
    P(\textrm{ex} \,i,j) = \textrm{min}\left[1,\exp\left(\gamma^{-1}\beta_i\left[U_i^{'}(\mathbf{R}_i) - U_i^{'}(\mathbf{R}_j)\right] + \gamma^{-1}\beta_j\left[U_j^{'}(\mathbf{R}_j) - U_j^{'}(\mathbf{R}_i)\right]\right)\right]
\end{equation}

\noindent
where $\beta_i = 1/ k_B T_i$ and $U_i^{'}$ is the potential energy of the $i$th replica including the EDS bias. Note that EDS will have different $\alpha$ for each replica because $\left<f\right>$ is a function of temperature. This exchange rule ensures that the resulting simulation will arrive at a canonical distribution without re-weighting. The result of this method is both an enhanced-sampled ensemble and a value for $\alpha$.

\section{Methods}
A short peptide GYG was simulated with EDS\cite{White2014} and PT-WTE\cite{Bonomi2010} to enhance sampling and improve fit to experimental NMR chemical shifts from Platzer et al.\cite{Platzer2014}. To compare the performance of EDS with enhanced sampling to other simulations, EDS with PT-WTE was compared to a single replica unbiased simulation, a single replica biased simulation, a parallel-tempering replica-exchange molecular dynamics (PT) simulation, and replica unbiased PT-WTE simulation.

0.008 mg/cm$^3$ GYG peptide in 10 mM NaCl counter-ions was first energy minimized then annealed and equilibrated in the NVT ensemble using the Canonical Sampling through Velocity Rescaling (CSVR) thermostat\cite{Bussi2007}. The CHARM27\cite{Pastor2011} force field and TIP3P\cite{Jorgensen1983} water model were used throughout. Electrostatic forces were calculated with the particle mesh Ewald method \cite{Essmann2007}, and dispersion forces were calculated with shifted Van der Waals potentials with a cutoff distance of 10 \AA. The covalent hydrogen bonds were constrained using the LINCS algorithm \cite{Hess1997} to enable a 2 fs time step. All simulations were performed in the Gromacs-5.1.4 simulation engine\cite{James2015}. To automatically run Gromacs commands in python, GromacsWrapper was used\cite{GromacsWrapper}. To generate an initial extended structure of GYG, the python package PeptideBuilder was used\cite{PeptideBuilder}. 

An equilibration is required for the PT-WTE method. 16 replicas of PT-WTE were tuned for 400 ps to find parameters that gave increased replica efficiency. The replica temperatures used for the 16 replica systems were 293, 299, 304, 310, 316, 323, 329, 336, 342, 349, 355, 363, 370, 377, 385, 392 K, as generated by the method of Nemoto et al. \cite{Nemoto}. The exchanges were attempted every 250 time steps. The WTE parameters were chosen following Barducci et al. \cite{Barducci2008} and are shown in Table 1. 

After the aforementioned short PT-WTE equilibration step, EDS was begun with the static bias from PT-WTE equilibration step. This combined PT-WTE with EDS had 16 replicas with the same set of temperatures as mentioned above and was 40 ns long. To control against the presence of either EDS bias, or PT-WTE bias, or both biases, 16 replica 40 ns PT simulations with and without EDS bias in the absence of WTE bias and 16 replica 40 ns PT-WTE simulations with and without EDS bias were run and as shown in Table 1. The large amount of time is not necessary for convergence of EDS, but was chosen to get good sampling statistics. The EDS collective variables biased were chemical shifts\cite{Platzer2014} calculated via the Plumed2 molecular simulation plugin\cite{Tribello2014}. The EDS parameters chosen are shown in Table 1.

An 8 replica PT-WTE simulation was also done with and without EDS. These simulations demonstrate how PT-WTE can use fewer replicas than PT and still achieve good efficiency. The parameters are in Table 1. The replicas ran at the following temperatures: 293, 306, 320, 335, 350, 366, 383, and 400 K. Similar to 16 replica PT-WTE, 8 replica PT-WTE initially consisted of 400 ps equilibration step, followed by a static production step of 40 ns. There were 3620 atoms in the simulation with 1197 tip3p water molecules and 38 peptide atoms. We used bias factor of 10 for 16 replica PT-WTE and 20 for 8 replica PT-WTE. Deighen et al. studied the effect of bias factor in 10 replica simulation that studied 39 amino acids long Trp-cage protein \cite{Deighan2012}. They found that increasing the bias factor from 10 to 24 decreased the the time it takes to converge to the reference FES\cite{Deighan2012}. Hence, we used larger bias factor for 16 replica PT-WTE compared to 8 replica PT-WTE.

\begin{table}

{\begin{tabular}{ ||cccccc||}
\hline

  \hline
  \multicolumn{6}{|c|}
  {\centering

  Overview of simulation systems and their parameters.}\\ [2ex]
    \hline
\hline
Simulation&EDS+&No EDS+&EDS+&No EDS+&Exp$^{(b)}$ \\ 
  Parameters&PT-WTE$^{(a)}$ & PT-WTE$^{(a)}$ &PT   &PT& \\
\hline
  Replicas&16&16&16&16&-\\
  Bias factor&10&10&-&-&-\\
  Hill width$\left(\frac{\textrm{kJ}}{\textrm{mol}}\right)$& 100&100&-&-&-\\
  Hill height $\left(\frac{\textrm{kJ}}{\textrm{mol}}\right)$&0.1&0.1&-&-&-\\
  Dim. Prop &0.2&-&0.2&-&-\\
  Range&0.01&-&0.01&-&-\\
  Time (ns)&40(0.4$^{(a)}$)&40(0.4$^{(a)}$)&40&40&-\\
\hline

  \hline
  
  \multicolumn{6}{||c||}{Results}\\
[2ex]
    \hline
 
  C$\beta$ $\delta$ (ppm)&38.52$\pm$2.35 &36.39$\pm$2.13 &38.48$\pm$1.84 & 39.06$\pm$2.89 & 38.60$\pm$0.08 \\ 

  C$\alpha$ $\delta$ (ppm)& 57.89$\pm$3.21 &59.38$\pm$ 2.35 &57.93$\pm$3.25 & 58.23$\pm$3.34 & 58.00$\pm$0.08 \\ 

  H $\delta$ (ppm)& 8.29$\pm$0.50 & 8.20$\pm$0.48 &8.20$\pm$0.46 &8.25$\pm$0.47&8.16$\pm$0.02  \\ 

  H$\alpha$ $\delta$ (ppm)& 4.58$\pm$0.51&  4.04$\pm$0.52& 4.54$\pm$0.42 & 4.61$\pm$0.55& 4.55$\pm$0.02 \\ 

  CO $\delta$ (ppm)& 176.28$\pm$1.30& 174.78$\pm$1.11&176.25$\pm$1.26 &175.34$\pm$1.64&176.30$\pm$0.08   \\ 
  N $\delta$ (ppm)&120.06$\pm$5.34  & 118.34$\pm$7.43&120.09$\pm$5.35& 119.88$\pm$6.58&120.1$\pm$0.08 \\

  Repl Efficiency (\%)  & 39(31$^{(a)}$) & 21(31$^{(a)}$) &22& 23 & - \\
  Aver Efficiency (\%)  & 41(40$^{(a)}$) & 28.2(40$^{(a)}$)&23 &25 & - \\
 
\end{tabular}}
{\begin{tabular}{ ||cccccc||}
\hline

\hline
  \multicolumn{6}{|c|}{Overview of simulation systems and their parameters.}\\ [2ex]
    \hline

Simulation&EDS+&No EDS+&EDS+&No EDS+&Exp$^b$ \\ 
  Parameters&PT-WTE&PT-WTE&No PT&No PT& \\
  \hline
  Replicas&8&8&1&1&-\\
  Bias factor&20&20&-&-&-\\
  Hill width$\left(\frac{\textrm{kJ}}{\textrm{mol}}\right)$&250&250&-&-&-\\
  Hill height $\left(\frac{\textrm{kJ}}{\textrm{mol}}\right)$&1&1&-&-&-\\
  Dim. Prop &0.2&-&0.2&-&-\\
  Range&0.001&-&0.01&-&-\\
  Time (ns)&40(0.4$^{(a)}$)&40(0.4$^{(a)}$)&40&40&-\\

  \hline
  \multicolumn{6}{|c|}{Results}\\[2ex]
    \hline

  C$\beta$ $\delta$ (ppm) &38.48$\pm$2.06&37.77$\pm$3.05&38.47$\pm$1.71&39.18$\pm$2.58& 38.60$\pm$0.08 \\ 

  C$\alpha$ $\delta$ (ppm) &57.94$\pm$3.35&58.91$\pm$3.14&57.94$\pm$3.06&57.99$\pm$3.33& 58.00$\pm$0.08 \\ 

  H $\delta$ (ppm)& 8.20$\pm$0.47&8.22$\pm$0.48&8.37$\pm$0.50&8.21$\pm$0.49&8.16$\pm$0.02  \\ 

  H$\alpha$ $\delta$ (ppm)&4.54$\pm$0.45&4.26$\pm$0.61&4.54$\pm$0.41&4.68$\pm$0.51& 4.55$\pm$0.02 \\ 

  CO $\delta$ (ppm)& 176.25$\pm$1.30 &175.00$\pm$1.55&176.30$\pm$1.14&175.40$\pm$1.54&176.30$\pm$0.08   \\ 
  N $\delta$ (ppm)&120.12$\pm$5.49 &117.90$\pm$7.20&120.01$\pm$4.79&119.43$\pm$5.84&120.1$\pm$0.08 \\

  Repl Efficiency (\%)  &55 (40$^{(a)}$)&43(39$^{(a)}$) &-&-& - \\
  Aver Efficiency (\%)  &   62 (51$^{(a)}$)&58(53$^{(a)}$) &-&  -&- \\ 
\hline
 

\end{tabular}}
\par
\bigskip
{\bf{Table 1}}. These are the computed averages during various simulations. 2.5$^{th}$ and 97.5$^{th}$ percentile  of the computed values are shown as well. The EDS and PT-WTE parameters are shown first as well as the simulation time. The chemical shifts are the cumulative averages. The replica-exchange efficiency is the percentage of attempted replica swaps that are successful for the least efficient replica. The average is the efficiency averaged over all replicas.

 {\bf{a}}-PT-WTE was equilibrated for 400 ps and then restarted without further addition of gaussian hills for the 40 ns simulation. RMSD due to Camshift\cite{Kohlhoff2009} for N,HN,HA,CA,CB, and C  are 3.01,0.56,0.28,1.3,1.36 and 1.38 ppm.
 
{\bf{b}}-Experimental values are from Platzer  et al \cite{Platzer2014}
\end{table}

\section{Results \& Discussion}

Table 1 provides an overview of the simulations run for this work. Multiple combinations of enhanced sampling methods with and without EDS were simulated. Each simulation ran for 40 ns. The 16 replica PT-WTE simulations were equilibrated for 400 ps to find the biasing potentials that increased the replica-exchange efficiency, prior to enabling EDS. This PT-WTE tune step was run long enough to ensure the efficiency exceeded 30\%. Parallel-tempering replica-exchange (PT) with 16 replicas has 23\% replica efficiency, as shown in Table 1. The replica efficiency is defined here as the minimum acceptance rate of configuration swapping between replicas. In other words, the lowest among the 16 replicas is taken to be the efficiency. This ensures that all replicas exchange at certain threshold rate prohibiting a different independent system of simulations that are exchanging only with replicas that have closer temperatures. The averages are also shown for the 40ns simulations. Replica-exchange of 20\% is the minimum for a well-sampled system\cite{James2015}, and thus 16 was the minimum number of replicas for PT.

The choice of PT-WTE simulation parameters achieved good exchange with the smaller, 8 replica system, despite spanning the same temperature range as the 16 replica systems (293K to 400K). Since 8 replica system had a different set of WTE parameters, EDS parameters were modified as well, which included the reduction of the speed of convergence of EDS (lower range parameter) and the increase the speed of convergence of the PT-WTE method (larger hill height)\cite{Bonomi2010}. The bias factor was correspondingly decreased to reduce the strength of bias more quickly. PT-WTE with 16 replicas had smaller replica efficiency than 8 replica PT-WTE in Table 1. This may be the result of a different set WTE and EDS parameters used in 8 replica and 16 replica simulations.

The EDS method moved the calculated backbone chemical shifts to the experimental values when applied as seen in Table 1. There is good agreement between the average chemical shifts in the EDS simulations and experiments with any of the enhanced sampling methods applied. Even with the slower parameters used in the 8 replica system, the EDS method still achieved the correct values.

\begin{figure}
\centering
{
\includegraphics[width=6.1cm,height=4.6cm]{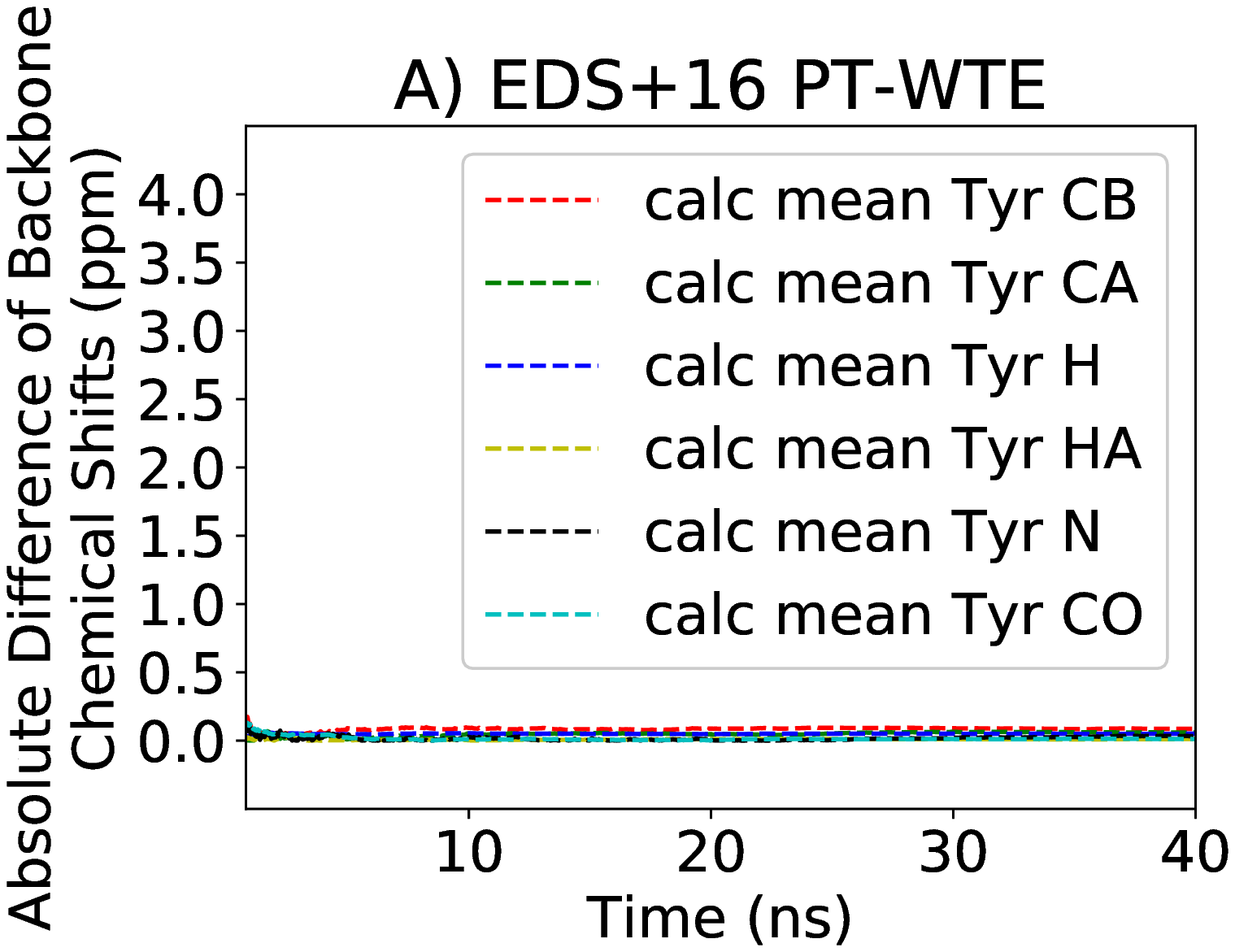}\label{fig:shiftsa}}
{
\includegraphics[width=6.1cm,height=4.6cm]{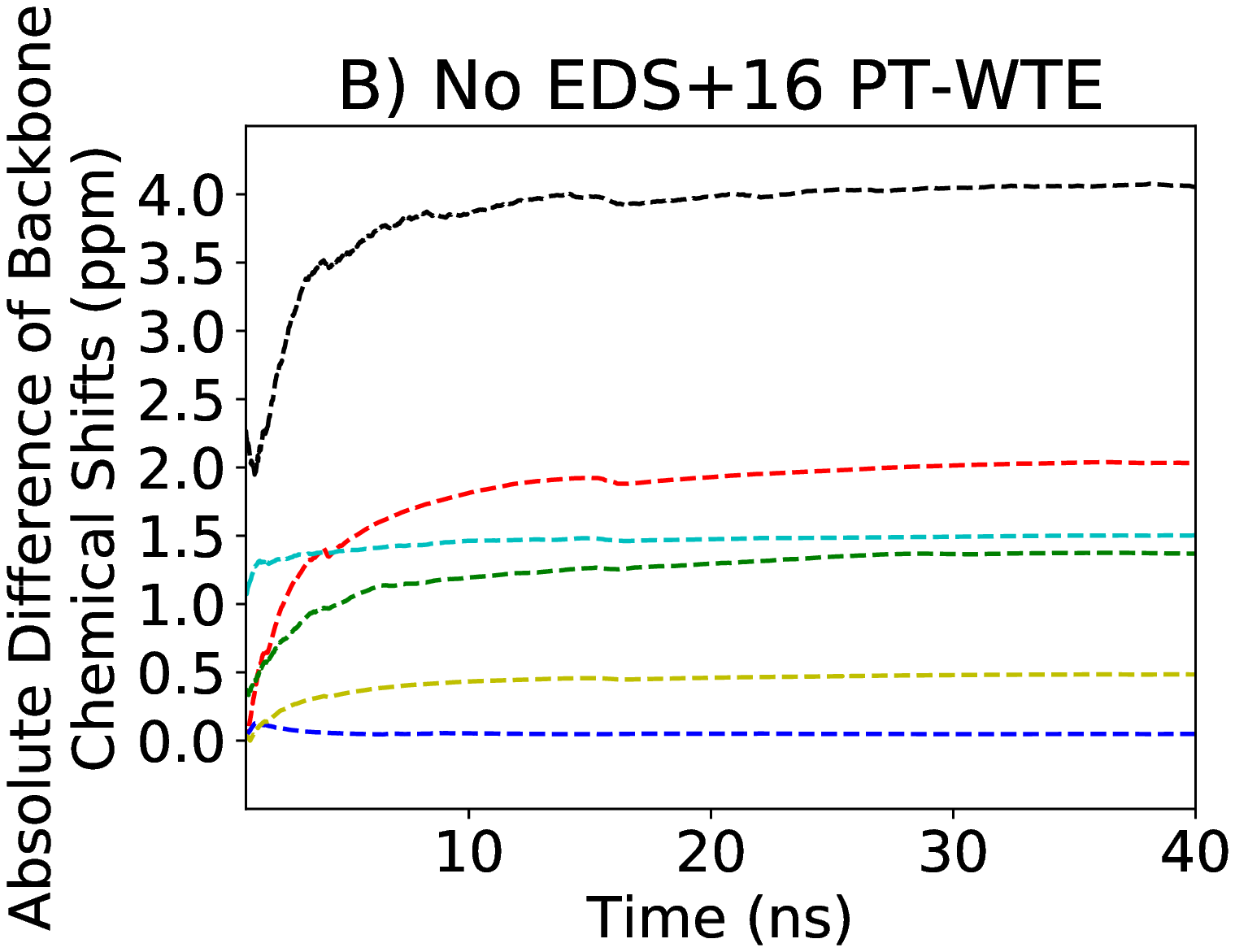}\label{fig:shiftsb}}%

\hspace{0mm}
{
\includegraphics[width=6.1cm,height=4.6cm]{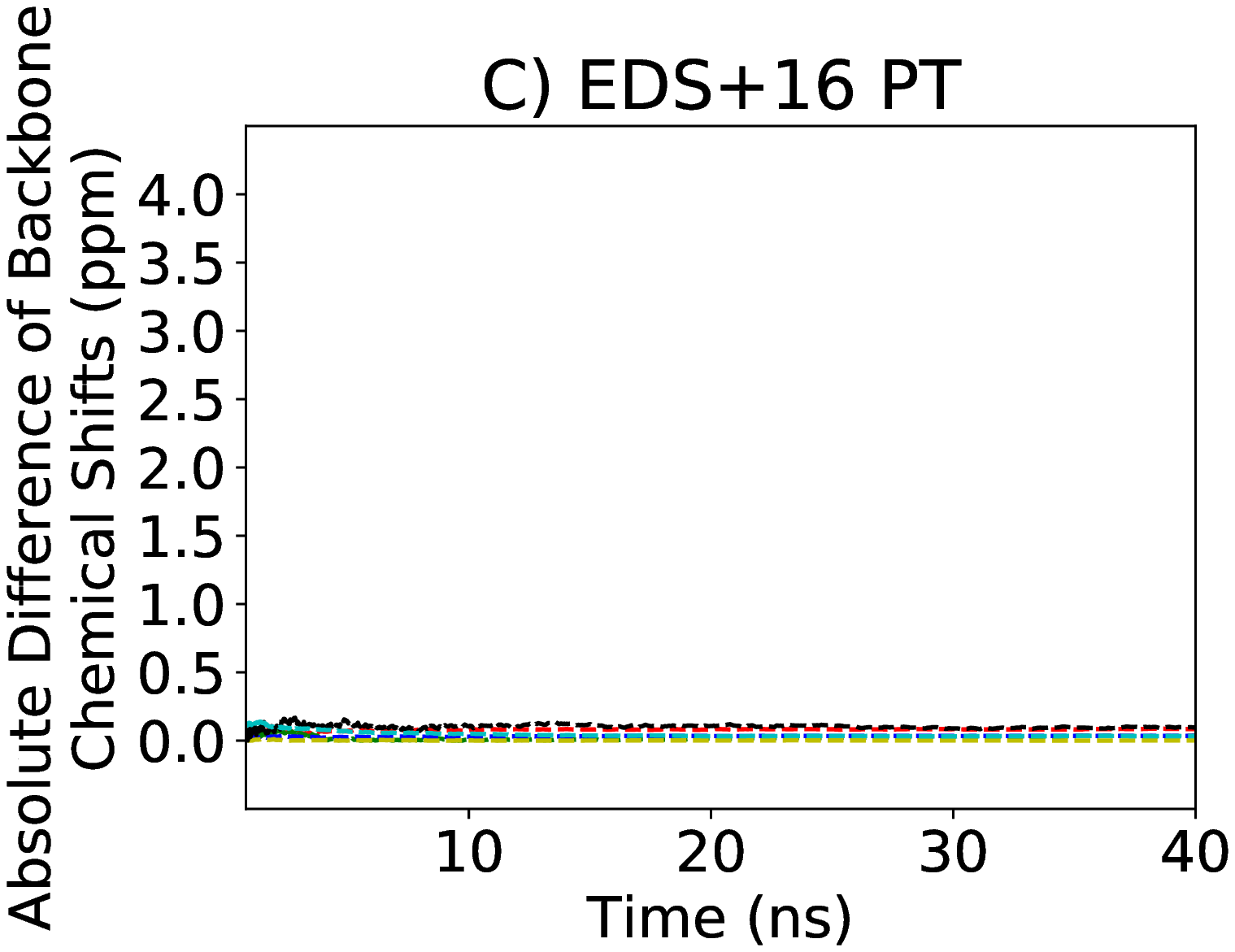}\label{fig:shiftsc}}
{
\includegraphics[width=6.1cm,height=4.6cm]{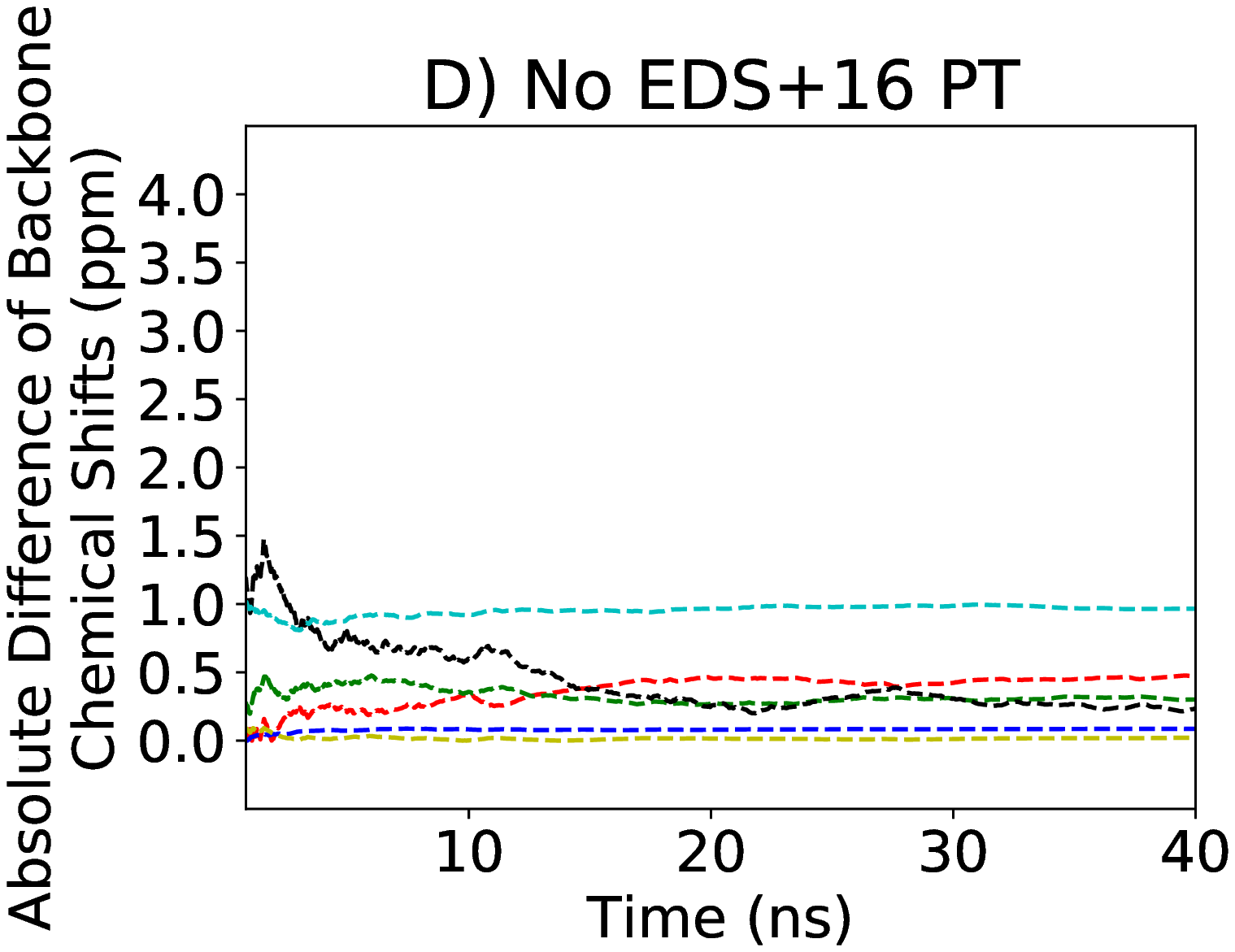}\label{fig:shiftsd}}%
\hspace{0mm}
{
\includegraphics[width=6.1cm,height=4.6cm]{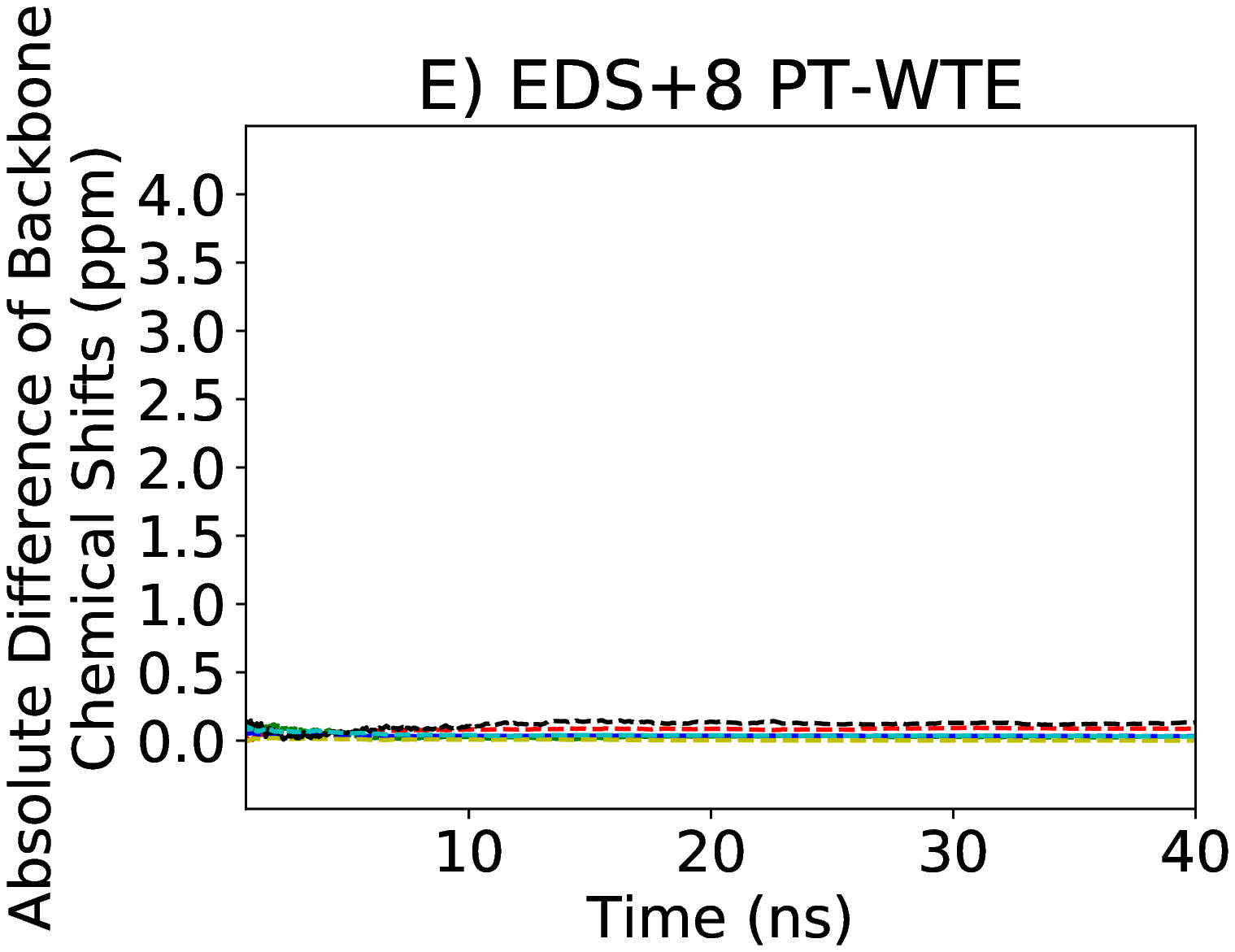}\label{fig:shiftse}}
{
\includegraphics[width=6.1cm,height=4.6cm]{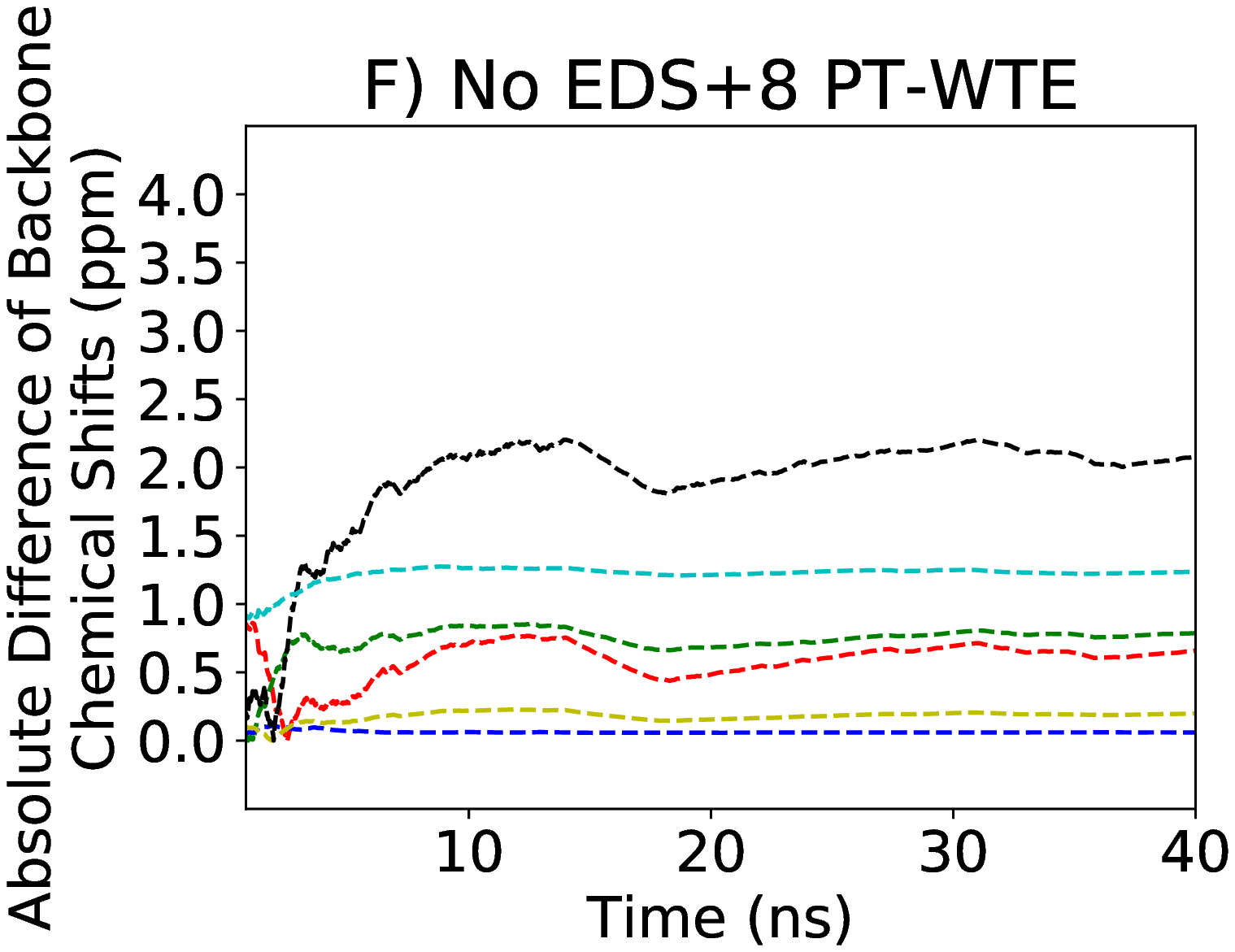}\label{fig:shiftsf}}%
\hspace{0mm}
{
\includegraphics[width=6.1cm,height=4.6cm]{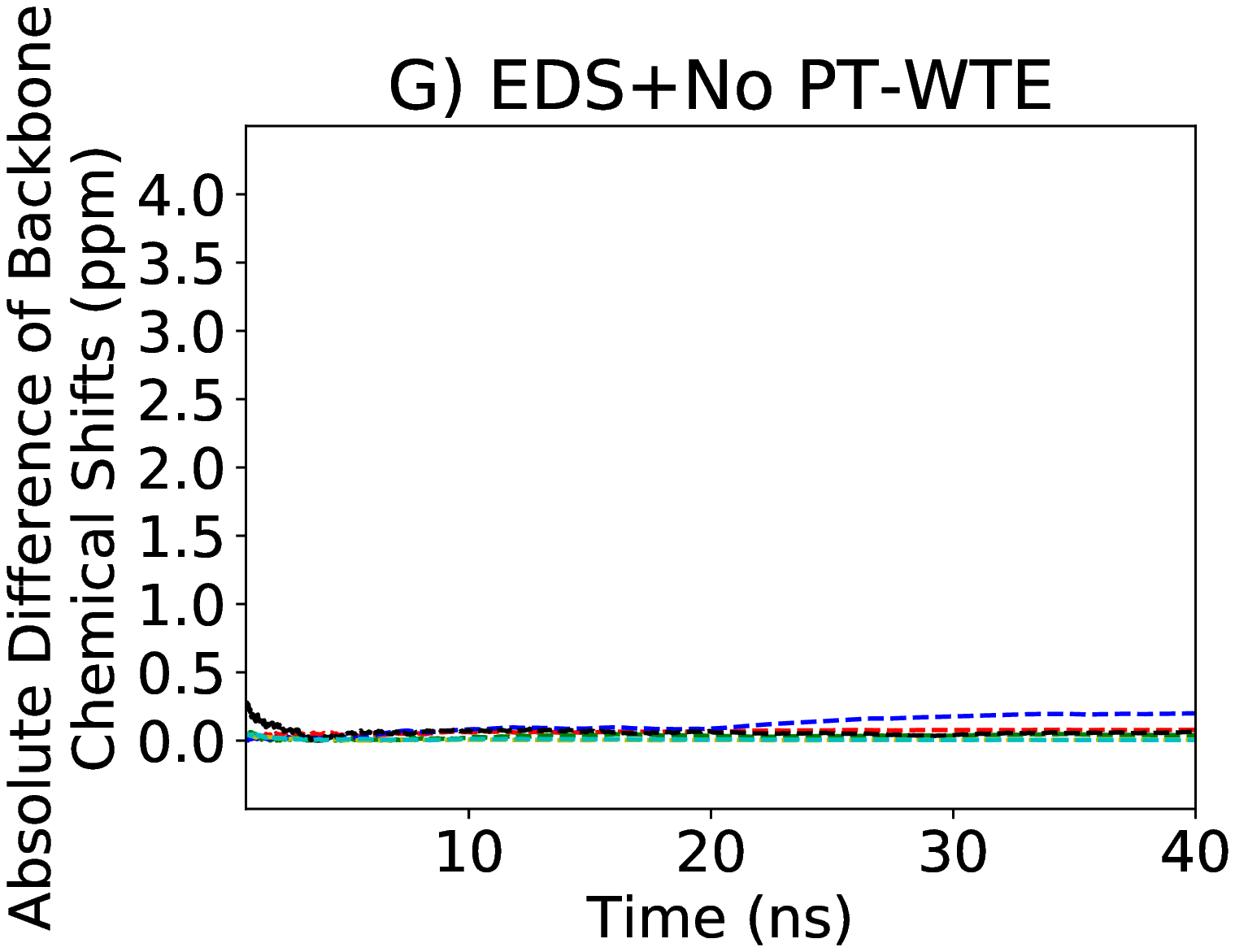}\label{fig:shiftsg}}
{
\includegraphics[width=6.1cm,height=4.6cm]{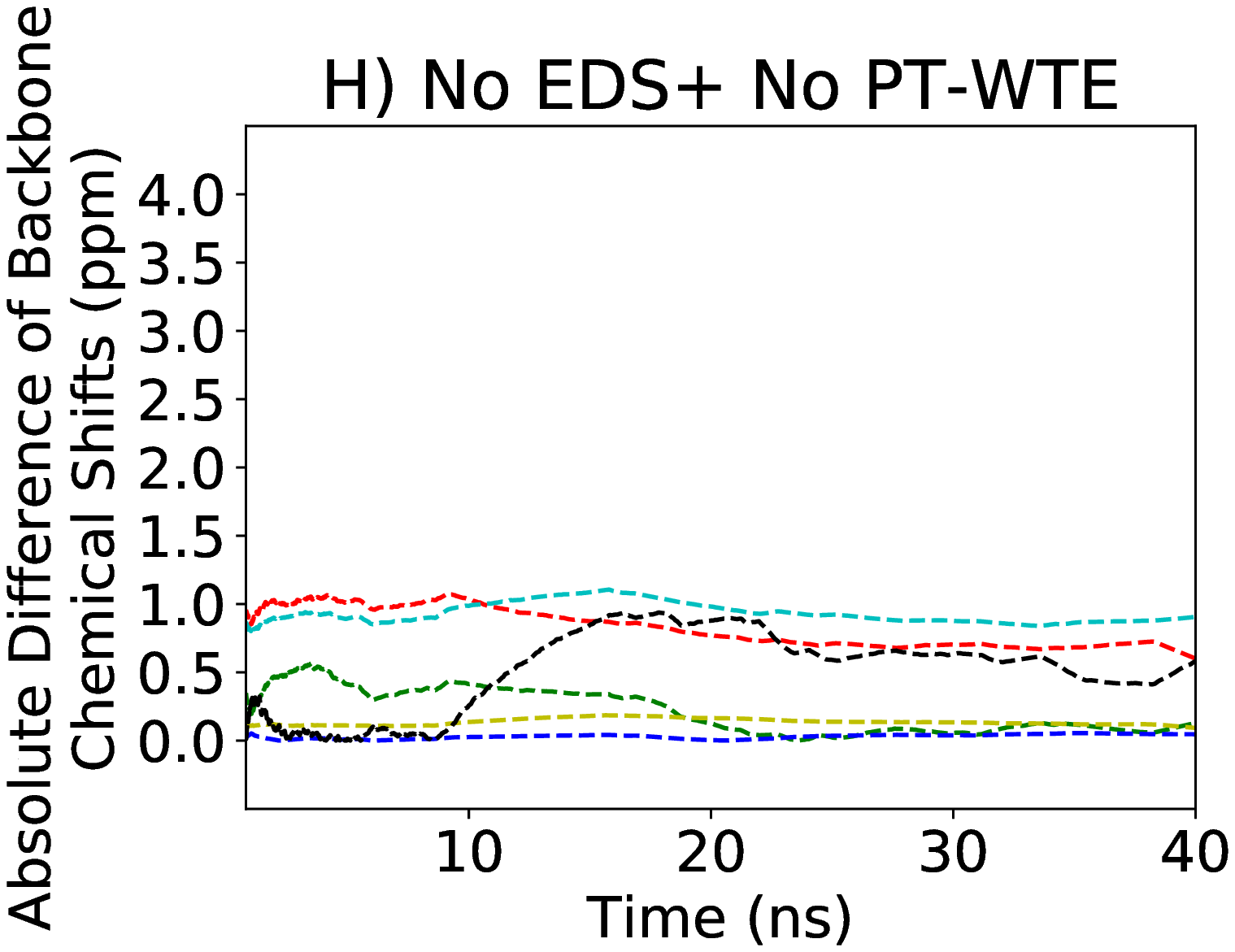}\label{fig:shiftsh}}%
\caption{Convergence of chemical shifts to the reference experimental value. Absolute difference between calculated and experimental chemical shifts. EDS was used with PT-WTE for enhanced sampling as shown in (a). (b) shows that without EDS. (c) EDS with 16 replica PT converges to experiments. The lack of EDS lowers the convergence (d). (e) displays with fewer replicas, EDS still converges to experiments.(f) Without EDS, convergence is lower.}
\label{fig:shifts}
\end{figure}

Figure~\ref{fig:shifts} shows the cumulative calculated mean backbone chemical shifts over time during 16 replica PT-WTE with EDS bias (a), 16 replica PT-WTE without EDS (b), 16 replica PT with EDS (c), 16 replica PT without EDS, 8 replica PT-WTE with EDS (e), 8 replica PT-WTE without EDS (f), single replica EDS (g), and a standard MD (h). Chemical shifts were modified to show the absolute difference between cumulative averages and the corresponding experimental chemical shifts. Therefore, exact match between simulation and experiment is indicated by the convergence of cumulative calculated mean chemical shifts to 0. As expected, EDS biased average chemical shifts converge to the reference value with 16 replica PT-WTE (Fig. 1a), with 16 replica PT (Fig. 1c), and with 8 replica PT-WTE (Fig. 1e). EDS does not make instantaneous values of biased chemical shifts converge with the reference value, but makes the ensemble average to approach the reference value as shown in fig. S1 of the Supporting Information. In all of the EDS cases (fig. 1a, 1c, 1e, and 1g), EDS converges within 1 ns. In all EDS bias simulations (fig. 1a, 1c, 1e, and 1g), average chemical shifts of the hydrogen connected to backbone nitrogen atom had the highest deviation from predicted value compared to other atoms. The Camshift root mean square deviation for H connected to N is 0.56 ppm, whereas Camshift predicts deviations to be highest for N with RMSD of 3.01 ppm in a 28 protein test set\cite{Kohlhoff2009}. During 40 ns simulation, the cumulative average backbone HN chemical shift was within 1 \% of the reference chemical shift. 
Since PT-WTE and PT swaps configurations between different temperature replicas, chemical shifts also change whenever the conformations change as indicated by the greater fluctuations between mean chemical shifts and reference values in PT-WTE and PT with EDS (Fig. 1a, 1c, and 1d ) simulation compared to single replica EDS (Fig. 1g). In the absence of EDS bias, average chemical shifts in various simulations, shown in figures 1b, 1d, 1e, and 1f, do not converge to 1 because the simulation is less accurate. When EDS was absent in both 16 replica PT-WTE (Fig. 1b) and 16 replica PT (Fig. 1d), average chemical shifts fluctuated more drastically in Fig. 1b compared to simulation without PT-WTE as shown in Fig. 1d. This is due to the better exploration of phase-space with the PT-WTE method. The convergence under a variety of enhanced sampling methods shows that EDS still works well when combined with enhanced sampling.

\begin{figure}[H]
\centering
\subfloat{ 
\includegraphics[scale=0.45]{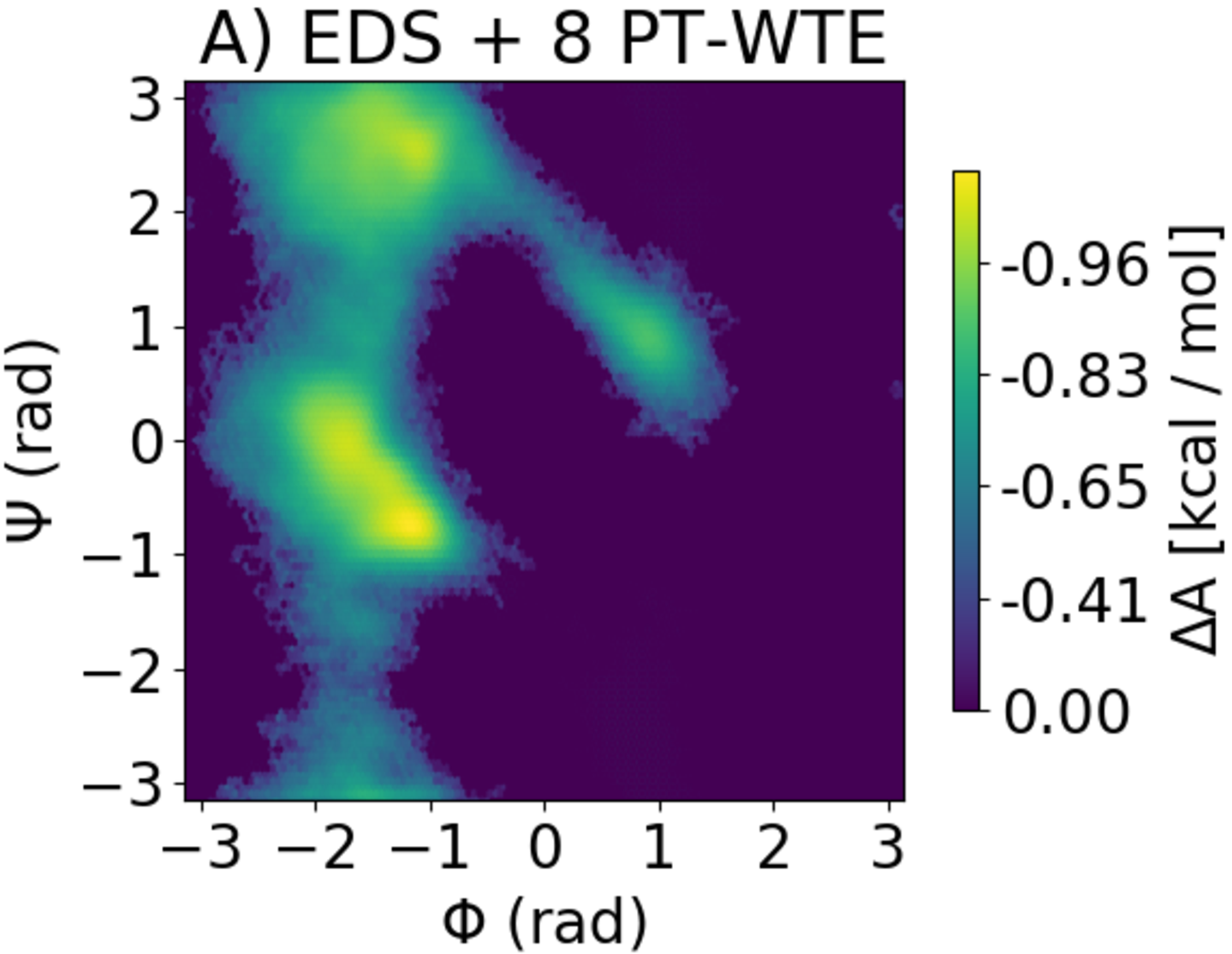}\label{fig:pmfa}}
\subfloat{
\includegraphics[scale=0.45]{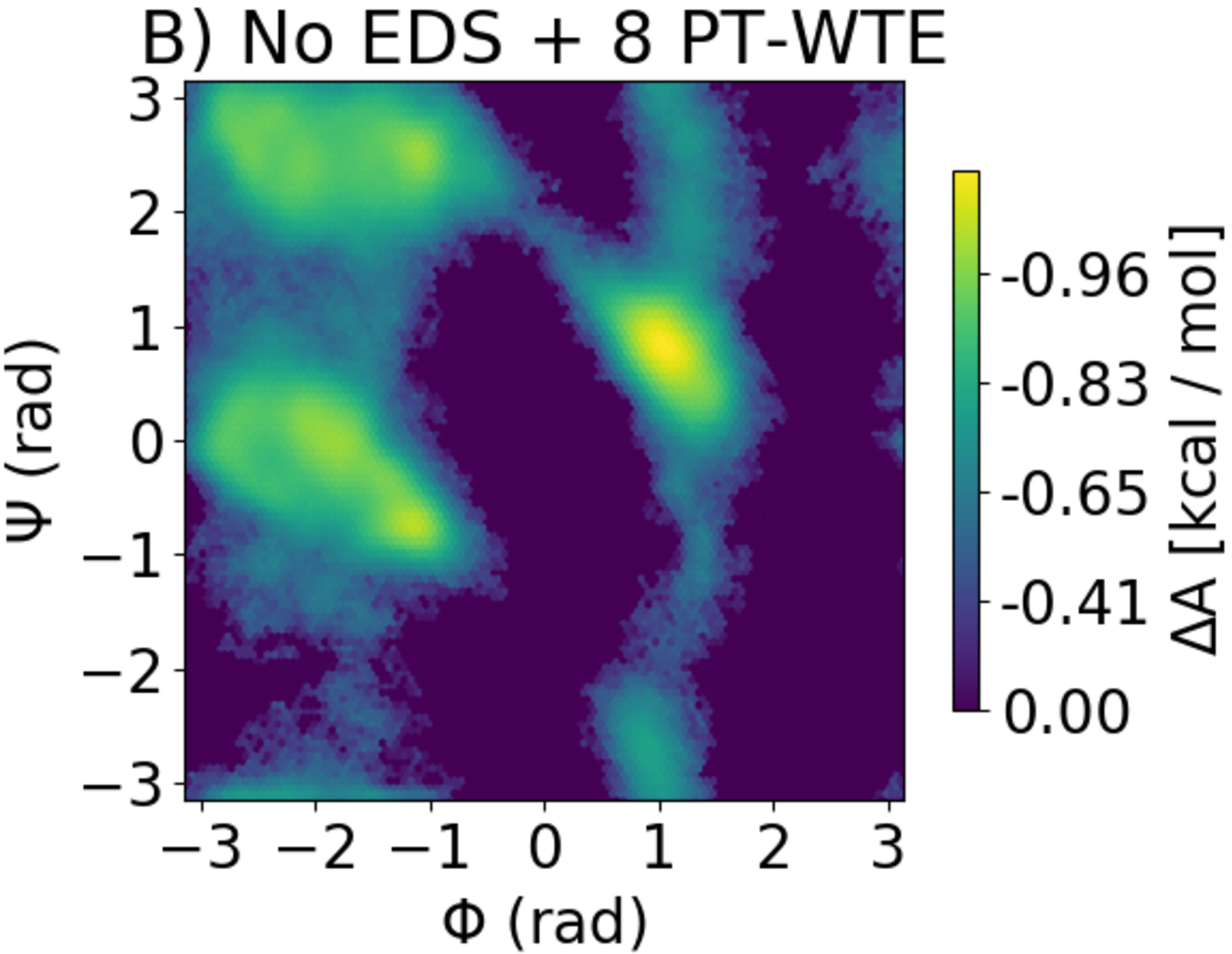}\label{fig:pmfb}}\hspace{0mm}%
\subfloat{
\includegraphics[scale=0.45]{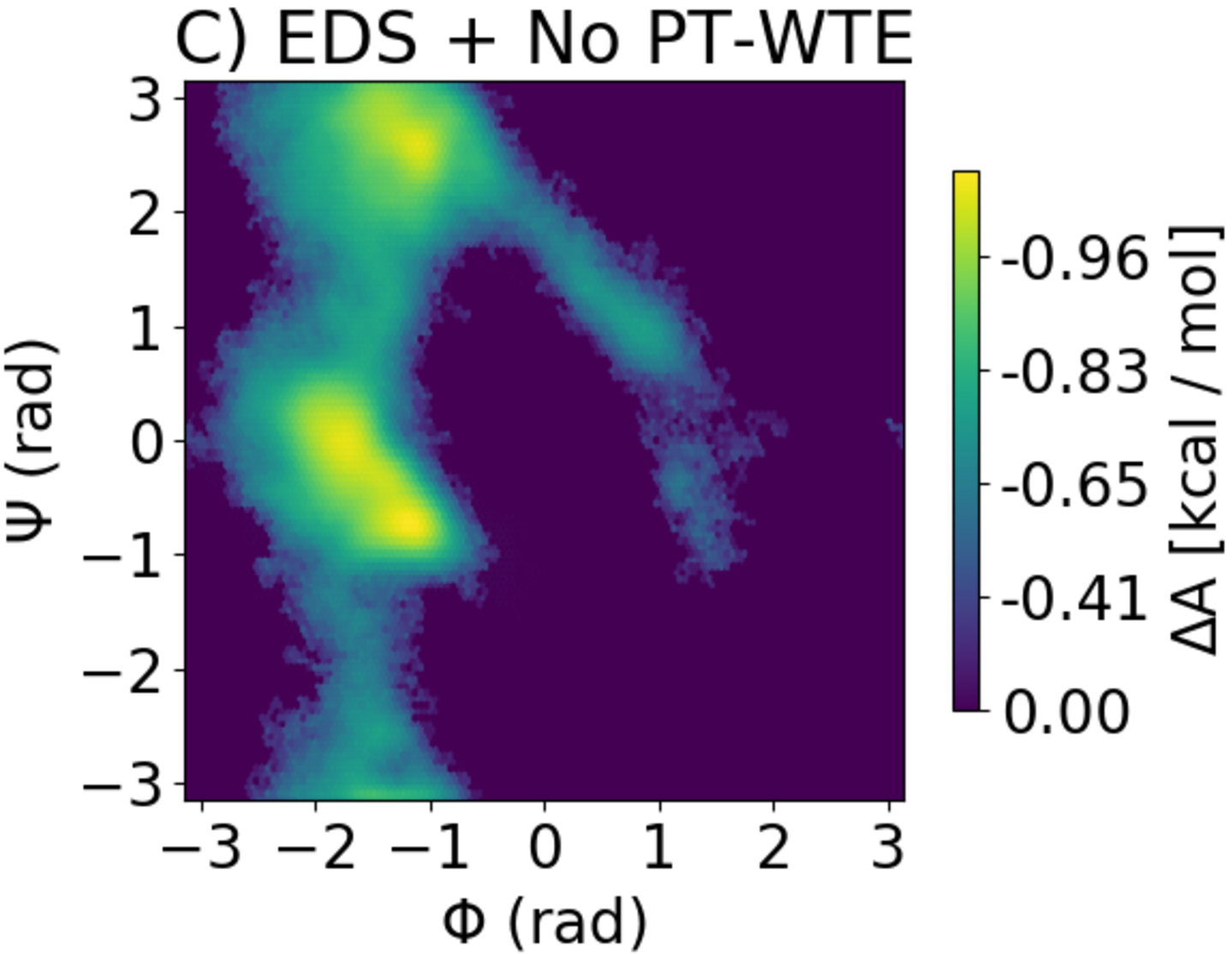}\label{fig:pmfc}}
\subfloat{
\includegraphics[scale=0.45]{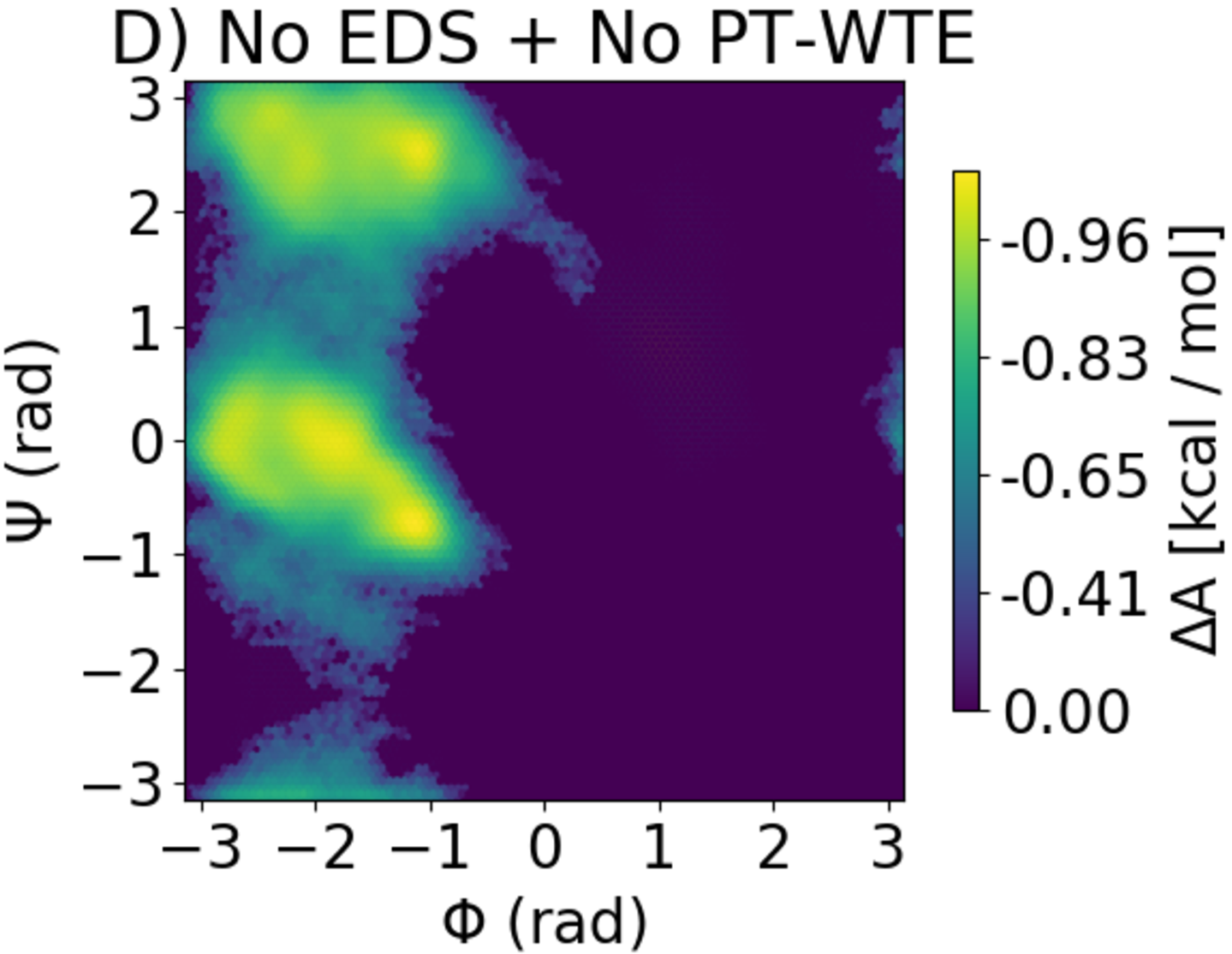}\label{fig:pmfd}}%
\hspace{0mm}
\caption{Free Energy Surface (FES) of $\Phi$-$\Psi$ dihedral angles of Y amino acid from GYG simulations. (a) shows the FES with EDS and PT-WTE so that backbone chemical shifts match experimental values and sampling is enhanced. (b) without EDS bias, PT-WTE explores $\Phi$ around 1.5 and $\Psi$ in -3.14,3.14 radians interval more, and finds a global minimum at about $\Phi$=1.5,$\Psi$=1 rad. (c) shows less sampling of global minimum at $\Phi$=1.5 radians and $\Psi$=1 radians,compared to a). (d) shows no  exploration of the global minimum at $\Phi$=1.5 and $\Psi$=1 rad and seems to be stuck on negative $\Phi$ angles. All FESs were generated from a 20 ns NVT simulation at 293K}
\label{fig:pmf}
\end{figure}

To see the effect of enhanced sampling and the effect of biased chemical shifts to other collective variables in a simulation, $\Phi$ and $\Psi$ dihedral angles were compared to regular unbiased MD and to Ramachandran plots from X-ray crystallography\cite{Ting2010}. The simulated free energy landscapes of Tyrosine in GYG peptide along $\Phi$ and $\Psi$ dihedral angles are  displayed in Fig.~\ref{fig:pmf}. Dihedral angles were computed during the simulations and their histograms were used to compute the FES. EDS predicted the same global minimum regardless of presence (Fig. ~\ref{fig:pmfa}) or absence of PT-WTE (Fig. ~\ref{fig:pmfc}) at $\Phi$=-1.2 and $\Psi$=-0.9. The free energy surface shows different global minimum with EDS because the free energy distribution is biased to satisfy experimental conditions. As shown in figures ~\ref{fig:pmfa} and ~\ref{fig:pmfc}, EDS simulations show the global minimum to be at $\Phi$=-1.2 and $\Psi$=-0.9 radians, different from the one observed by enhanced sampling alone (Fig. ~\ref{fig:pmfb}). Enhanced sampling technique found a different global minimum at $\Phi$=1 and $\Psi$=1, in the absence of EDS than in the presence of EDS. Also, PT-WTE without EDS enabled the exploration of a region at $\Phi$=1 and $\Psi$ in interval [-3.14,3.14] radians in figure ~\ref{fig:pmfb} which is similar to Tyrosine Ramachandran map from Vitalini et al\cite{Vitalini2016a}; neither of EDS alone (Fig.~\ref{fig:pmfc}), EDS with PT-WTE (Fig.~\ref{fig:pmfa}) and regular MD (Fig.~\ref{fig:pmfd}) could sample around that region continuously. Although EDS in the presence of PT-WTE (Fig.~\ref{fig:pmfa}), did not show aberrant behavior from absence of PT-WTE in figure ~\ref{fig:pmfc}, it showed improved sampling of global minimum observed in PT-WTE alone at $\Phi$ around 1 radians and at $\Psi$=1 radians compared to the absence of PT-WTE (Fig.~\ref{fig:pmfc}) and regular MD (Fig. ~\ref{fig:pmfd}). Overall, PT-WTE improved sampling and EDS modified the FES to achieve agreement with the NMR chemical shifts.

Ramachandran plots from Ting et al.\cite{Ting2010} of Tyrosine in the sequence GYG were compared to our results. The Ting et al.\cite{Ting2010} results are derived from a bioinformatics analysis of X-ray crystallographic data of whole proteins. Tyrosine dihedral angle plot of Tyrosine in YG peptide from Ting et al.\cite{Ting2010}, was compared to our results, since YG was closest GYG. Ting et al, observed three regions of minimum in Tyrosine: at about ($\Phi$=-1.0,$\Psi$=2.8, which was the global minimum), ($\Phi$=-1.0,$\Psi$=[-1,1] interval), and ($\Phi$=1,$\Psi$=1) rad. Out of all the FES for simulations that are shown, Fig.~\ref{fig:pmfa} was closest to experimentally observed FES by Ting et al.\cite{Ting2010}, because, local minimum at ($\Phi$=1,$\Psi$=1) was predicted correctly and minima at ($\Phi$=-1.0,$\Psi$=2.8) and ($\Phi$=-1.0,$\Psi$=[-1,1] interval) were close relative to each other in Ting et al.\cite{Ting2010} and in Fig.~\ref{fig:pmfa}. Ting et al.\cite{Ting2010} also reported the global minima to be on the negative $\Phi$ angle axis. Without EDS, enhanced sampling alone could not label the experimentally observed global minimum as global minimum at $\Phi$=-1 and $\Psi$=2.8 in Fig.~\ref{fig:pmfb}. $\Psi$ global minimum was around zero for both Ting et al.\cite{Ting2010} and EDS results (Fig.~\ref{fig:pmfa} and ~\ref{fig:pmfc}). A local minimum at around $\Phi$=1 and $\Psi$=1 radians reported by Ting et al.\cite{Ting2010}, was also observed as a minimum Fig.~\ref{fig:pmfa}, ~\ref{fig:pmfb}, and ~\ref{fig:pmfc}. This local minimum was not predicted by standard MD at all within 20 ns, as seen in Fig.~\ref{fig:pmfd}. EDS did not show drastic deviation in unbiased structural properties, such as a dihedral angle of Tyrosine. Indeed dihedral angles in EDS simulations with or without enhanced sampling showed similar results as X-ray crystallography did.

Another major question about disagreement between simulations and experiments is whether the underlying cause is lack of sampling or inaccuracy of the force-field/system. The 8 replica PT-WTE without EDS shows that even with good sampling, the simulation matches neither the chemical shifts nor the FES from the EDS simulation, as shown in Fig.~\ref{fig:pmfb}.  This is not necessarily due to a deficiency in the CHARMM force field, but instead could be due to a difference in concentration, ions, and termini between the simulations and the work of Platzer et al. \cite{Platzer2014}, from which the chemical shifts were measured.

\section{Conclusion}

Simultaneous enhanced sampling and experiment directed simulation (EDS) was demonstrated on the GYG peptide in explicit solvent. EDS improved the accuracy of the simulation by minimally biasing the backbone chemical shifts of the peptide to match experimental data from Platzer et al\cite{Platzer2014}. Single replica simulations with and without EDS bias ran for comparable time. Both parallel-tempering replica-exchange (PT) and parallel-tempering well-tempered ensemble (PT-WTE) were used with EDS to improve sampling. Compared to absence of enhanced sampling in a control MD, system has significantly benefited from enhanced sampling. PT and PT-WTE provided improved sampling and did not interfere with EDS, with EDS converging within 1 ns. On the same computer architecture, 16 replica PT-WTE simulations and 16 replica PT with and without EDS completed after approximately the same period of wall-clock time. PT-WTE with 8 replicas spanning 293K to 400K was also demonstrated with EDS, showing that the PT-WTE method is able to reduce the number of replicas required relative to PT while not jeopardizing enhanced sampling. We have demonstrated that PT-WTE and PT can be applied with EDS simultaneously without interfering with each other in this system. We will exploit this new method on larger system of proteins and peptides, to take advantage of both enhanced sampling as well as biasing with experiments. 

\section{Acknowledgements}
Support for DB Amirkulova provided by a  Hopeman scholarship from University of Rochester. Thanks to Rainier Barret for feedback and discussion of manuscript. We are grateful for resources and services provided by Center for Integrated Research Computing at University of Rochester.
\bibliographystyle{ieeetr}

\bibliography{dialanine}

\end{document}


\markboth{Dilnoza B Amirkulova and Andrew D White}
{Combining Enhanced Sampling with Experiment Directed Simulation of the GYG peptide}

%

%

\title{Supporting Information for Combining Enhanced Sampling with Experiment Directed Simulation of the GYG peptide}

\author{Dilnoza B Amirkulova,}
\author{Andrew D White}

\renewcommand{\thefigure}{S\arabic{figure}}
\maketitle

\begin{figure}[H]
\centering
\subfloat{ 
\includegraphics[scale=0.6]{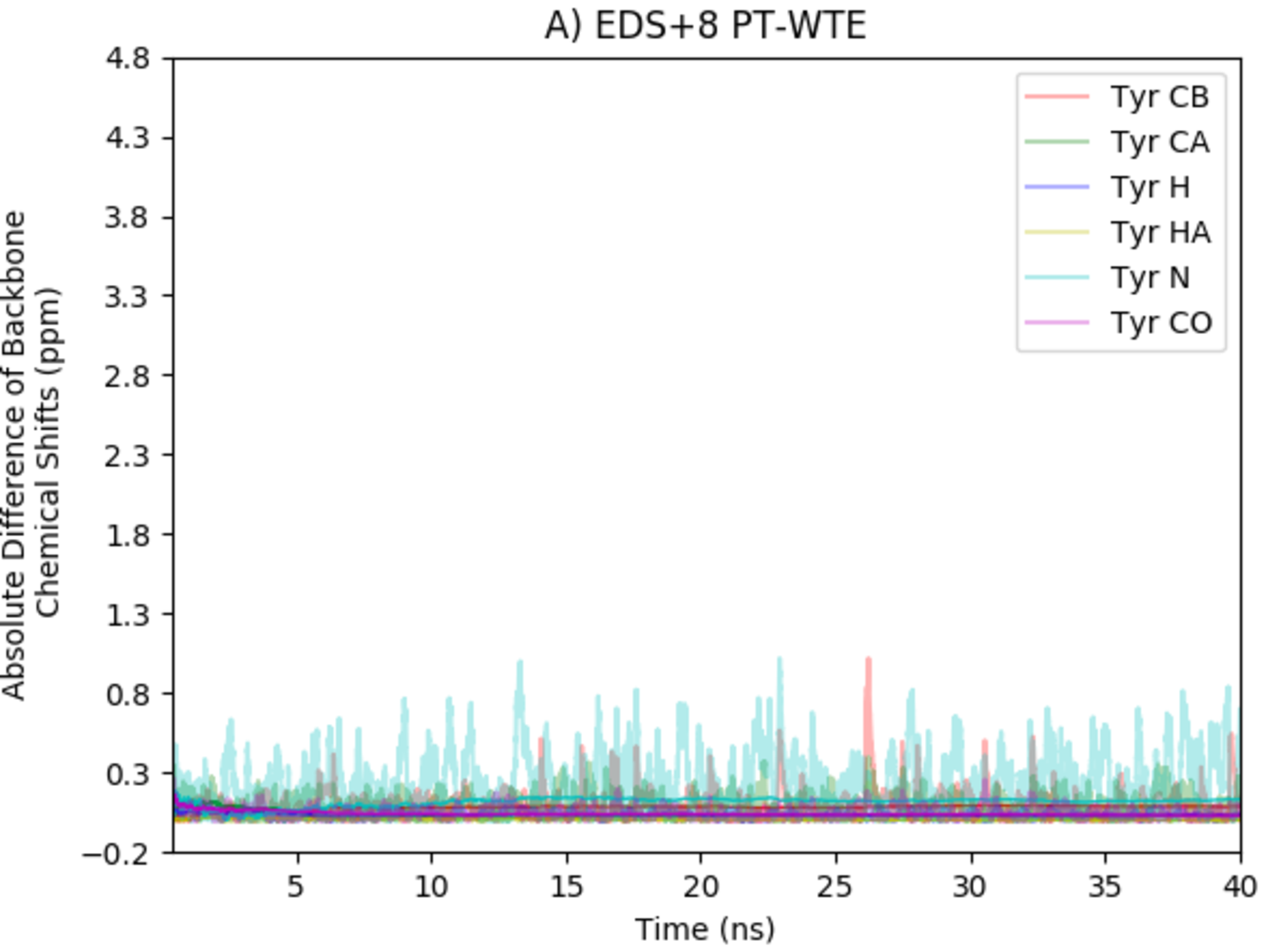}\label{fig:}}\\
\hspace{0mm}
\subfloat{ 
\includegraphics[scale=0.6]{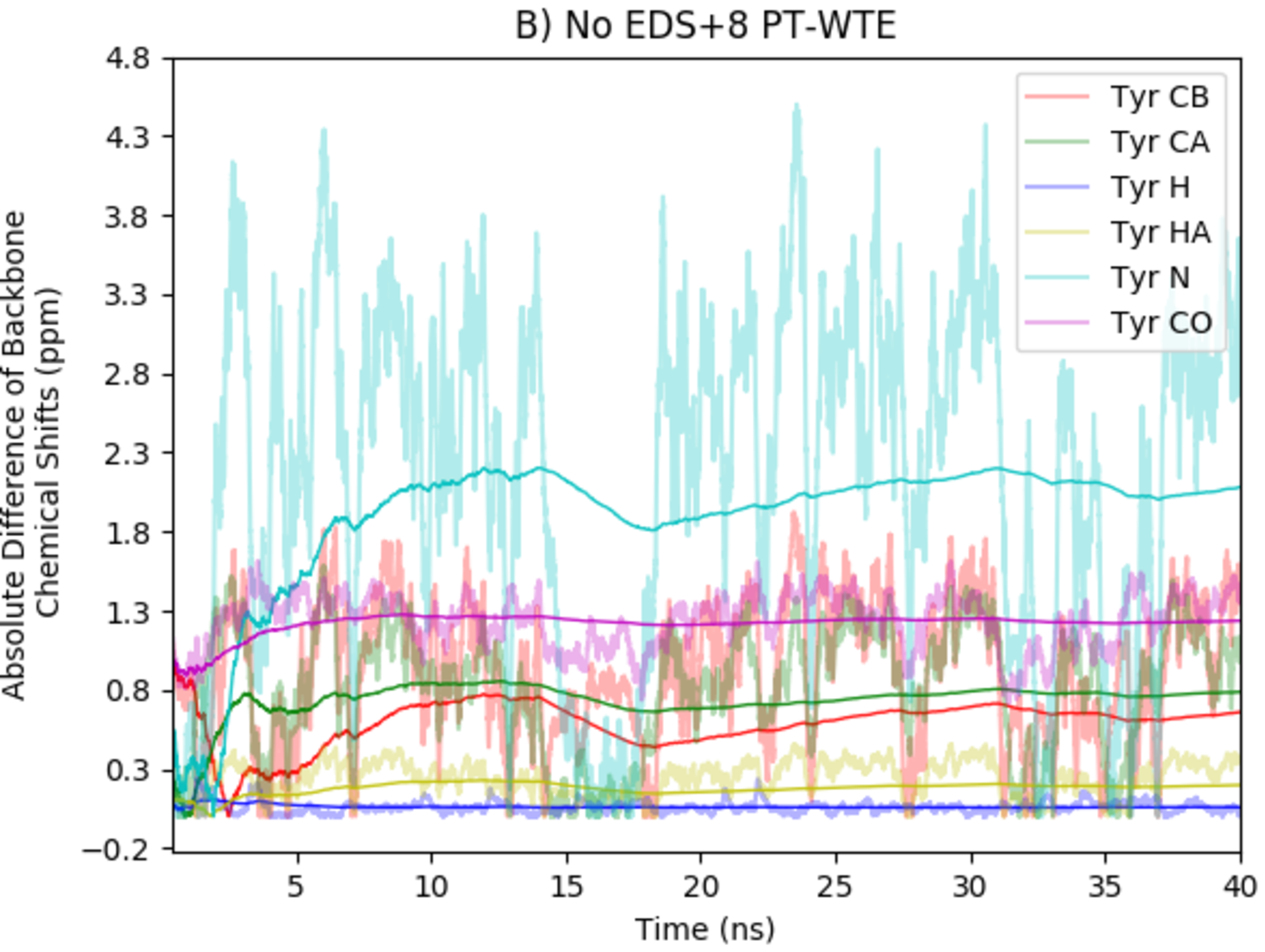}\label{fig:}}
\hspace{0mm}
\caption{Window Moving Average of Chemical Shifts Compared with Cumulative Average in 8 replica PT-WTE with (A) and without EDS (B). Cumulative Average of Chemical shifts is converging to 0, whereas moving window average of last 0.4 ns window is shown in less opaque shade of the same color. EDS does not try to match instantaneous values of chemical shifts to reference, but rather makes the cumulative average match the reference. Both moving averages and cumulative averages were subtracted by the corresponding backbone chemical shifts from experiments, which EDS is biasing to match. Perfect match is when cumulative average approaches 0. }
\label{fig:mover_aver}
\end{figure}